\documentclass{llncs}
\usepackage{t1enc}

\begin{document}
\mainmatter              
\title{{\LARGE ICOOOLPS'2006}\\ECOOP Workshop on {\underline
  I}mplementation,  {\underline C}ompilation,   {\underline
  O}ptimization of  {\underline O}bject-{\underline O}riented
  {\underline L}anguages, {\underline P}rograms and {\underline S}ystems} 
\titlerunning{ICOOOLPS'2006}  
%
\author{Roland Ducournau\inst{1} 
\and Etienne Gagnon\inst{2}
\and Chandra Krintz\inst{3}
\and Philippe Mulet\inst{4}
\and Jan Vitek\inst{5}
\and Olivier Zendra\inst{6}
}

\institute{LIRMM, France 
\and  UQAM, Canada
\and  UCSB, USA	
\and  IBM, France
\and  Purdue University, USA
\and  INRIA-LORIA, France	
}

%
%
%

\maketitle              

\begin{abstract}
ICOOOLPS'2006 was the first edition of ECOOP-ICOOOLPS workshop.
It intended to bring researchers and practitioners
both from academia and industry together, with a spirit of openness,
to try and identify and begin to address the numerous and very varied
issues of optimization. 
This succeeded, as can be seen from the papers, the attendance and the
liveliness of the discussions that took place during and after the
workshop, not to mention a few new cooperations or postdoctoral
contracts. 
The 22 talented people from different groups who participated were
unanimous to appreciate this first edition and recommend that ICOOOLPS
be continued next year. 
A community is thus beginning to form, and should be reinforced by a
second edition next year, with all the improvements this first edition
made emerge.
\end{abstract}
%

\section{Objectives and call for papers}

Object-oriented languages are pervasive and play a significant role in
computer science and engineering life and sometime appear as
ubiquitous and completely mature. However, despite a large number of
works, there is still a clear need for solutions for efficient
implementation and compilation of OO languages in various application
domains ranging from embedded and real-time systems to desktop
systems. 

The ICOOOLPS workshop thus aims to address this crucial issue of
optimization in OO languages, programs and systems. It intends to do
so by bringing together researchers and practitioners working in the
field of object-oriented languages implementation and
optimization. Its main goals are identifying fundamental bases and key
current issues pertaining to the efficient implementation, compilation
and optimization of OO languages, and outlining future challenges and
research directions. 

\smallskip

Topics of interest for ICOOOLPS include but are not limited to:
\begin{itemize}
    \item  implementation of fundamental OOL features:
       \begin{itemize}
          \item inheritance (object layout, late binding, subtype test...)
          \item genericity (parametric types)
          \item memory management
       \end{itemize}
    \item  runtime systems:
       \begin{itemize}
          \item compilers
          \item linkers
          \item virtual machines
       \end{itemize}
    \item  optimizations:
       \begin{itemize}
          \item static and dynamic analyses
          \item adaptive virtual machines
       \end{itemize}
    \item  resource constraints:
       \begin{itemize}
          \item real-time systems
          \item embedded systems (space, low power)...
       \end{itemize}
    \item  relevant choices and tradeoffs:
       \begin{itemize}
          \item constant time vs. non-constant time mechanisms
          \item separate compilation vs. global compilation
          \item dynamic loading vs. global linking
          \item dynamic checking vs. proof-carrying code
       \end{itemize}
\end{itemize}

This workshop tries to identify
fundamental bases and key current issues pertaining to the efficient
implementation and compilation of OO languages, in order to spread
them further amongst the various computing systems. It is also
intended to extend this synthesis to encompass future challenges and
research directions in the field of OO languages implementation and
optimization. 

Finally, this workshop is intended to become a recurrent one. Thus,
the organization (most relevant format and hottest topics) of this
workshop future occurrences will be adapted by the organizers and
attendees according to the main outcome of this workshop discussions.

In order to have a solid basis on which the discussions could be based
and to keep them focused, each prospective participant was 
required to submit either a short paper describing ongoing work or a
position paper describing an open issue, likely solutions, drawbacks
of current solutions or alternative solutions to well known problems. 
Papers had to be written in English and their final version could not
exceed 8 pages in LNCS style.

\section{Organizers}

\medskip

\begin{tabbing}
\hspace*{5mmm}\=\hspace*{2cm}\=\hspace{2cm}\=\kill
{\bf Olivier ZENDRA (chair)}, \>\>\>INRIA-LORIA, Nancy, France.\\
 \>Email:   \> {\tt olivier.zendra@loria.fr}\\
 \>Web:     \> {\tt http://wwW.loria.fr/\homedir zendra}\\
 \>Address: \> INRIA / LORIA\\
 \>         \> 615 Rue du Jardin Botanique\\
 \>	    \> BP 101\\
 \>	    \> 54602 Villers-Lès-Nancy Cedex, FRANCE\\
\end{tabbing}
\vspace{-5mm}
	Olivier Zendra is a full-time permanent computer science
	researcher at INRIA~/~LORIA, in Nancy, France. His research
	topics cover compilation, optimization and automatic memory
	management. He worked on the compilation and optimization of
	object-oriented languages and was one of the two people who
	created and implemented SmartEiffel, The GNU Eiffel Compiler
	(at the time SmallEiffel). His current research application
	domains are compilation, memory management and embedded
	systems, with a specific focus on low energy.

\begin{tabbing}
\hspace*{5mmm}\=\hspace*{2cm}\=\hspace{3cm}\=\kill
{\bf Roland DUCOURNAU (co-chair)}, \>\>\>~~~~~~~ LIRMM, Montpellier, France.\\
 \>Email:   \> {\tt ducour@lirmm.fr}\\
 \>Web:     \> {\tt http://www.lirmm.fr/\homedir ducour}\\
 \>Address: \> LIRMM,\\
 \>         \> 161, rue Ada\\
 \>	    \> 34392 Montpellier Cedex 5, FRANCE \\
\end{tabbing}
\vspace{-5mm}
	Roland Ducournau is Professor of Computer Science at the
	University of Montpellier. In the late 80s, while with Sema
	Group, he designed and developed the YAFOOL language, based on
	frames and prototypes and dedicated to knowledge based
	systems. His research topics focuses on class specialization
	and inheritance, especially multiple inheritance. His recent
	works are dedicated to implementation of OO languages. 

\medskip

\begin{tabbing}
\hspace*{5mmm}\=\hspace*{2cm}\=\hspace{3cm}\=\kill
{\bf Etienne GAGNON}, \>\>\>UQAM, Montréal, Québec, Canada.\\
 \>Email:   \> {\tt egagnon@sablevm.org}\\
 \>Web:     \> {\tt http://www.info2.uqam.ca/\homedir egagnon}\\
 \>Address: \> Département d'informatique\\
 \>         \> UQAM \\
 \>	    \> Case postale 8888, succursale Centre-ville \\
 \>	    \> Montréal (Québec) Canada / H3C 3P8\\
\end{tabbing}
\vspace{-5mm}
	Etienne Gagnon is a Professor of Computer Science at
	Université du Québec à Montréal (UQAM) since 2001. Etienne has
	developed the SableVM portable research virtual machine for
	Java, and the SableCC compiler framework generator. His
	research topics include language design, memory management,
	synchronization, verification, portability, and efficient
	interpretation techniques in virtual machines. 

\medskip

\begin{tabbing}
\hspace*{5mmm}\=\hspace*{2cm}\=\hspace{3cm}\=\kill
{\bf Chandra KRINTZ}, \>\>\>UC Santa Barbara, CA, USA.\\
 \>Email:   \> {\tt ckrintz@cs.ucsb.edu}\\
 \>Web:     \> {\tt http://www.cs.ucsb.edu/\homedir ckrintz}\\
 \>Address: \> University of California\\
 \>         \> Engineering I, Rm. 1121\\
 \>	    \> Department of Computer Science\\
 \>	    \> Santa Barbara, CA 93106-5110, USA\\
\end{tabbing}
\vspace{-5mm}

	Chandra Krintz is an Assistant Professor at the University of
	California, Santa Barbara (UCSB); she joined the UCSB faculty
	in 2001. Chandra's research interests include automatic and
	adaptive compiler and virtual runtime techniques for
	object-oriented languages that improve performance and
	increase battery life. In particular, her work focuses on
	exploiting repeating patterns in the time-varying behavior of
	underlying resources, applications, and workloads to guide
	dynamic optimization and specialization of program and system
	components. 

\medskip

\begin{tabbing}
\hspace*{5mmm}\=\hspace*{2cm}\=\hspace{3cm}\=\kill
{\bf Philippe MULET}, \>\>\>IBM, Saint-Nazaire, France.\\
 \>Email:   \> {\tt philippe\_mulet@fr.ibm.com}\\
 \>Address: \> IBM France - Paris Laboratory\\
 \>         \> 69, rue de la Vecquerie\\
 \>	    \> 44600 Saint-Nazaire, France\\
\end{tabbing}
\vspace{-5mm}

	Philippe Mulet is the lead for the Java Development Tooling
	(JDT) Eclipse subproject, working at IBM since 1996; he is
	currently located in Saint-Nazaire (France). In late 1990s,
	Philippe was responsible for the compiler and codeassist tools
	in IBM Java Integrated Development Environments (IDEs):
	VisualAge for Java standard and micro editions. Philippe then
	became in charge of the Java infrastructure for the Eclipse
	platform, and more recently of the entire Java tooling for
	Eclipse. Philippe is a member of the Eclipse Project
	PMC. Philippe is also a member of the expert group on compiler
	API (JSR199), representing IBM. His main interests are in
	compilation, performance, scalability and meta-level
	architectures.

\medskip

\begin{tabbing}
\hspace*{5mmm}\=\hspace*{2cm}\=\hspace{3cm}\=\kill
{\bf Jan VITEK}, \>\>\>Purdue Univ., West Lafayette, IN, USA.\\
 \>Email:   \> {\tt jv@cs.purdue.edu}\\
 \>Web:     \> {\tt http://www.cs.purdue.edu/homes/jv}\\
 \>Address: \> Dept. of Computer Sciences\\
 \>         \> Purdue University\\
 \>	    \> West Lafayette, IN 47907, USA\\
\end{tabbing}
\vspace{-5mm}

	Jan Vitek is an Associate Professor in Computer Science at
	Purdue University. He leads the Secure Software Systems
	lab. He obtained his PhD from the University of Geneva in
	1999, and a MSc from the University of Victoria in
	1995. Prof. Vitek research interests include programming
	language, virtual machines, mobile code, software engineering
	and information security.

\section{Participants}

ICOOOLPS attendance was limited to 30 people.
In addition, as mentioned in the call for paper, only people who were
giving a talk were initially allowed to attend ICOOOLPS. 
However, since on-site there were a lot of other people interested in
the workshop, the rules were relaxed to match the demand.

Finally, 22 people from 8 countries attended this first edition of
ICOOOLPS, filling the allocated room, as detailed in the following table: 

\bigskip 

\hspace{-17mm}\begin{tabular}{|l|l|l|l|l|}
\hline
First name & Name & Affiliation & Country & Email \\ 
\hline
\hline
Daniel & Benquides & EMN - Nantes & France & {\tt lbenquid@emn.fr} \\
Rhodes & Brown	    &  Univ. of Victoria & Canada & {\tt rhodesb@cs.uvic.ca} \\
Roland & Ducournau &	Univ. of Montpellier &	France & {\tt ducour@lirmm.fr} \\
Andres & Fortier &  LIFIA (UNLP) & Argentina & {\tt andres@lifia.info.unlp.edu.ar} \\
Etienne & Gagnon &  UQAM & Canada & {\tt egagnon@sablevm.org} \\
Olivier & Gruber &  IBM Research & France & {\tt ogruber@us.ibm.com} \\
Elisa & Gonzales Bax &	VUB & Belgium & {\tt egonzale@vub.ac.be} \\
Teresa & Higuera & UCM & Spain & {\tt mthiguer@dacya.ucm.es} \\
Yann & Hodique & USTL Lille 1 & France & {\tt hodique@lifl.fr} \\
Richard & Jones & Univ. of Kent & UK & {\tt R.E.Jones@kent.ac.uk} \\
Susanne & Jucknath & TU Berlin & Germany & {\tt susannej@cs.tu-berlin.de} \\
Eric & Jul & DIKU & Denmark & {\tt eric@diku.dk} \\
Chandra & Krintz & UC Santa Barbara & USA & {\tt ckrintz@cs.ucsb.edu} \\
Paul & McGregor & Goldman Sachs & USA & {\tt paul.regtech.mcgregor@gs.com} \\
Philippe & Mulet & IBM Rational Software & France & {\tt philippe\_mulet@fr.ibm.com} \\
Marco & Pistoia & IBM Watson Research &	USA & {\tt pistoia@us.ibm.com} \\
Jean & Privat &	  Univ. of Montpellier & France & {\tt privat@lirmm.fr} \\
Guillaume & Salagnac &	Verimag lab. & France & {\tt Guillaume.Salagnac@imag.fr} \\
Christophe & Rippert &	Verimag lab. & France & {\tt Christophe.Rippert@imag.fr} \\
Jan & Vitek & Purdue Univ. & USA & {\tt v@cs.purdue.edu} \\
Hiroshi & Yamauchi & Purdue Univ. & USA & {\tt yamauchi@cs.purdue.edu} \\
Olivier & Zendra & INRIA-LORIA &  France & {\tt Olivier.Zendra@loria.fr} \\ 
\hline
\end{tabular}

\section{Contributions}

All the papers and presentations are available from the ICOOOLPS web
site at {\tt http://icooolps.loria.fr}.

\subsection{Real-time and embedded systems}
\label{sec:contrib-RTES}

This session clustered papers and questions related to
real-time and/or embedded systems.

\bigskip

In ``Java for Hard Real-Time'', Jan Vitek presented the numerous
challenges caused by trying to put Java, a high-level, expressive
object-oriented language, in systems that require hard real-time
guarantees (such as avionics).
He detailed OVM (Open Virtual Machine), developed at Purdue Univ.

In ``Can small and open embedded systems benefit from escape analysis ?''
Gilles Grimaud, Yann Hodique and Isabelle Simplot-Rey explained how a
commonly known  technique, escape analysis, can be used in small
constrained embedded systems to improve time through a better memory
management, at low cost.

In ``Memory and compiler optimizations for low-power in embedded systems''
Olivier Zendra aimed at raising awareness about low-power and
low-energy issues in embedded systems among the object-oriented and
languages communities.
He showed how mostly known time- or size-optimization techniques can
be and should observed from a different point of view, namely energy.
He surveyed a number of solutions and outlined remaining challenges.

\bigskip

Based on the papers, presentations and discussions in  this session,
several trends clearly show. 

First, the ever increasing importance of embedded systems, whether they
are real-time of not, in software research.

Second, it could be argued (and has in the past) that, in such highly
constrained systems, the powerful features and expressiveness of
object-oriented languages and their compiler are too expensive to be
relied on. 
However, a trend can be seen in research that tries to bring these
features to smaller and smaller systems, trying to bridge a gap.
Hence, ``object-oriented'' and ``embedded'' are no longer opposite 
terms, but on the contrary form together a very active and promising
research area.  

Finally, new challenges (power, energy...) emerge, that require either
the proper integration of known techniques, or the development of new
ones. 
As such, being able to take into account low-level (hardware)
features at high level (OO, JVM...) appear quite challenging but offer
a high potential.

It is however of course always very challenging to both be able to
increase the level of abstraction and at the same time get a finer,
lower-level understanding of the application.

\subsection{Memory management}

This session grouped papers whose main topic was memory
management. 

\bigskip

``Efficient Region-Based Memory Management for Resource-limited
Real-Time Embedded Systems'', by Chaker Nakhli, Christopher Rippert,
Guillaume Salagnac and Sergio Yovine, presents a static algorithm to
make dynamic region-based memory allocations for Java applications
that have real-time constraints.
M. Teresa Higuera-Toledano addresses close issues, aiming at
``Improving the Scoped Memory-Region GC of Real-Time Java''.

This confirms the growing importance of real-time for
object-oriented languages in general, and more specifically Java, with
the RTSJ (Real-Time Specification for Java).
This is additional evidence for the trend we mentioned in section
\ref{sec:contrib-RTES} towards bringing high expressiveness, easy to
use languages in smaller and/or more constrained systems 

Richard Jones and Chris Ryder argued, in ``Garbage Collection Should
be Lifetime Aware'', that the efficiency of garbage collectors can be
improved by making them more aware of actual objects lifetimes in the
mutator. 
Indeed, even current generational garbage collectors generally observe
the mutator with a rather coarse view, and do not provide enough
flexibility when clustering objects according to their expected
lifetimes.
This is an area where the potential gain in performance is quite
considerable. 

This presentation and the following discussions where quite refreshing
and confirm that even in a rather technical and well explored domain,
new ideas can emerge that have both high potential and are relatively
easy to grasp, especially when explained in a metaphorical way.

Finally, in ``Enabling Efficient and Adaptive Specialization of
Object-Oriented, Garbage Collected Programs'', Chandra Krintz defended
code optimizations (specialization) which are aggressively and
speculatively performed and can be, if the need arises, invalidated on
the fly, through OSR (On Stack Replacement).

Here again, we can spot the trend that was mentioned during the
discussion for session \ref{sec:contrib-RTES} and tends to bridge the
gap between hardware and software.
Indeed, the presented technique bear some similarities with what
processors do in hardware, with speculative execution and invalidation.

\bigskip

All the above mentioned papers and discussions make it clear that
memory management is an area where a lot of progress can be made, be
it in small or large strides.
Memory management is furthermore an area which has an important impact
over program speed.
In addition to speed, memory management can also affect very
significantly energy usage, as discussed during session
\ref{sec:contrib-RTES}.
Memory-targeted optimizations should thus always be taken into
account when trying to reach higher performance.

\subsection{Optimization}

This session was devoted to papers and questions related known or very
specific optimizations.

\bigskip 

In ``OO specific redundancy elimination techniques'', Rhodes Brown and
Nigel Horspool advocated a holistic view of optimization where not
only one but in fact the whole set of program properties are taken
into account together, without forgetting their mutual interactions. 
They presented how annotations could be used, in conjunction with
static and dynamic invariants, to improve program performance. 

This echoes a relatively novel trend in object-oriented program
optimization, that tries to analysis not only one specific
optimization, but optimization composition or sequences.

``Performing loop optimizations offline'', by Hiroshi Yamauchi, shows
how the overhead of some loop optimizations can be removed from
execution time by performing them offline.
This is especially important in the context of system that require a
high level of responsiveness.

This shows how even very specific and potentially common optimizations
can be reconsidered in the light of new constraints, for example
those of the real-time systems that were already mentioned in session
\ref{sec:contrib-RTES}.

\bigskip

Here again, as in session \ref{sec:contrib-RTES}, we see that not
optimization work tends to evolve.
First, they more and more focus not only on one criterion which is
often program speed, but also integrate other criteria, such as
responsiveness, that correspond more to new current computing systems
with tight real-time and/or space constraints.
Second, larger sets of optimizations tend to be considered together,
as optimization sequences or compositions, to better encompass the
complexity inherent to real life systems and the various interactions
that can take place when optimizing.

\subsection{Abstraction and frameworks}

This last session aimed at regrouping papers and talks about higher
level or broader points of view for optimization.

\bigskip

In ``Efficient Separate Compilation of OO languages'' Jean Privat,
Floréal Morandat and Roland Ducournau present a scheme to reconcile
separate and global compilation, hence global optimization.
They detail their practical and implemented solution in the context of
an object-oriented language.

This is truly another example of research work trying to successfully 
bridge a gap: the gap between separate compilation, which commonly
used in industry, and global compilation, that brings the best
optimization results.

``Java Framework for Runtime Modules'' by Olivier Gruber and Richard
Hall is a paper that takes a broad view of optimization.
It proposes a framework to more easily build modules and reuse
components and that could be integrated in the Java Runtime
Environment. 

By bringing this discussion to the workshop, the authors clearly
enlarged to scope of the discussions and tried to connect the
optimization and software engineering communities.
This kind of openness is quite useful in a workshop so as to foster
slightly unusual cooperation and work.

Finally, ``The Importance of Abstraction in Efficient OO
implementations'' by Eric Jul, made a case for clearly and strictly
separating the abstraction (language level) and the concrete
(implementation) level.
This indeed gives more freedom to the implementer, hence more
possibilities for optimizations, while the language user does not
have to worry about low-level details but only about the semantics of
the program.

Here, we see that bridging the gap between what is expressed and what
is implemented is important, but should not be left to the developer. 
That's the compiler's job, or rather the compiler implementers' job.
Eric's position is thus quite important to remind us not to pollute the
high level with too many low-level details.
Of course, one question that remains open is how to properly abstract
things, especially low-level, possible hardware, details.

\bigskip

This session was interesting in that it made the workshop participant
not forget a high-level, software engineering oriented point of view
and the related issues.
Indeed, there is always a risk that, being focused on one specific
optimization, the researcher forgets the larger picture.
Considering issues at a high level, with abstraction, may avoids
getting swamped in details.
Reuse of optimizations, like reuse of modules, is a requirement to
evolve from software optimizations as a craftsmanship to software
optimizations as an industrial process.
Of course, quite some work remains to be done before we're there, but
it a goal worth aiming at.

\section{Closing debates}

The presentation sessions finished later than scheduled.
As a consequence, the discussion time that was planned at the end of
the workshop was shorter than initially expected.
This may have somehow limited the discussions.

This is one of the points that shall be improved in future occurrences
of ICOOOLPS (see section \ref{sec:conclusions-perspectives}).

\bigskip

The discussion session was very spontaneous, with attendees being
encouraged to bring their favorite topic, main itch, etc.
From their summary emerge two main treads.

\subsection{``Written down in code vs. inferred''}

"The user knows" what is intended and what is going on in an
application. Thus, it seems to make sense to have the developer
{\em annotate the code with meta information}, that can then be used by the
compiler to optimize. 

However the code --- especially for libraries --- can be reused in a
different context. 
Would the annotations remain valid ? 
This seems to call for {\em context-dependent annotations}. 

But what is ``the context'' when you write 10\% and reuse 90\% ? 

``Annotation-guided optimization'' looks quite appealing. 
However, the analyses done by the compiler have to be performed
anyway, whether annotations are present or not. 
What should the compiler do if annotations appear to be 
{\em contradictory} with what it infers ? 

Relying on developer annotations puts a burden on her/him. 
But we all know there are good developers and not-so-good ones, with a
majority in the second category, so is it {\em realistic} ? 

There are similarities between the user-software interface and the
hardware-software interface: interactions are needed, as well as
information passing (both ways). 
For example, feedback to the user is very useful, so that s/he can
improve her/his coding.  

The developer knows the application, but should not have to worry
about the underlying OS, hardware, etc. 
Annotations thus should make it possible to express {\em what the
 developer wants, not how to do it}.

\bigskip

A lot of interest was expressed in this long debate with many
attendees involved.
 
Annotations by the developer seem appealing but their nature is an
issue. 
A lot depends on the developer level, so how far can annotations be
trusted ?  

This discussion thread certainly is worth digging deeper into during
the next edition of ICOOOLPS.

\subsection{``Do threads make sense ?''}

Isn't the threading model fundamentally flawed, that is
inappropriate/problematic for object-oriented design and
implementation ?
Indeed, threads in Java are build on top of an existing, independent model. 
They thus seem poorly integrated. 

See Hans-J. Boehm, "Threads Cannot Be Implemented as a Library" - PLDI
2006 and Edward A. Lee, "The Problem with Threads" - IEEE Computer,
May 2006. 

\bigskip 

This topic, however, did not spark much debate, maybe because of lack
of time.  

\section{Conclusion and perspectives}
\label{sec:conclusions-perspectives}

This first edition of ICOOOLPS was able to reach one its goals:
bringing together people from various horizons, in an open way, so as
to foster new, original and fruitful discussions and exchanges
pertaining to optimizations.
The presence of people from 8 different countries, from academia  and
industry, researcher as well as practioners, is in itself a success.
The fact that more people that expected showed up is another.

Thanks to the skills of the speakers and active participation of the
attendants, the discussions were lively, open-minded and allowed good
exchanges. 
Identifying the mains challenges for optimization is not that easy
though. 
Indeed, as emerged more clearly during ICOOOLPS, optimizations for
object-oriented languages come in variety of contexts with very
different constraints (embedded, real-time, dynamic, legacy...).
The optimizations criteria considered, thus the goal, also tend to
differ a lot: speed, size, memory footprint, more recently energy...
In addition, all these have to be tackled keeping in my higher-level,
software engineering-oriented issues, such as modularity,
composability, reusability, ease of use...

Some trends can however be sketched.
Optimizations tend to encompass more and more target criteria
(multi-criteria rather than single criterion), such as energy and
speed, or memory footprint and responsiveness.
Multiple optimizations tend to be evaluated in conjunction, as
sequences of optimizations, rather than in an isolated way.
Separating semantics and implementation is crucial, for
expressiveness, ease of use and the possibility to perform
optimizations at compile level.
However, it appears at the same time necessary to be able to better
take into account the actual execution of a program when optimizing,
that is better take into account the behavior of the software and the
hardware as well.

Large challenges thus remain, and should be addressed by the
community.
That's what ICOOOLPS intend to do in its next editions.

\bigskip

Indeed, the perspectives for the ECOOP-ICOOOLPS workshop appear quite
bright. 
One of the questions was whether this workshop should be
pursued in the next years, and with which periodicity.
The answer was unanimously positive: attendees are in favor of
continuing the workshop next year with a yearly periodicity.

Overall satisfaction is thus quite high for this very first edition of
the workshop. 

A few ways to improve ICOOOLPS emerged during the workshop and should
be taken into account in 2007: 

\begin{itemize}
    \item Presentations should be significantly shorter, to save time
    for longer discussions. The later should take place during the
    sessions, for example one session comprising 3 talks lasting 5 to
    10 minutes each, plus a 30 to 60 minutes  discussion. 
    \item More time could also be allotted for discussions at the
    very end of the workshop.
    \item Session report drafts should be written during a session
    (papers and talks) and maybe briefly discussed at the end of each
    session (not after the workshop).
    \item Attendees could be given the possibility to submit
    (written) questions to paper presenters before the workshop
    itself. This would give a starting base for discussions, or at the
    very least the question ``session'' at the end of each talk. 
    \item  The workshop could be open to anyone, not only
    authors/speakers. This year indeed, although no call for
    participation had been issued after the call for paper was closed,
    because the workshop was for presenters only, many more people
    asked to be admitted in. Since the aim of an ECOOP workshop is to
    foster discussions and exchanges, refusing interested people would
    have been a bad idea. Having everyone (not only authors) present
    themselves and their work in a few minutes would be an added value.
    \item A larger room is necessary. 15 attendants were expected but
    22 came, so the room was very crowded, which made it 
    difficult for some attendants to properly see the presentation
    slides. 
\end{itemize}

\section{Related work}

In order to provide a fixed point for ICOOOLPS related matters, the
web site for the workshop is maintained at {\tt
http://icooolps.loria.fr}. 
All the papers and presentations done for ICOOLPS'2006 are freely
available there.

%

\nocite{*} 

\bibliographystyle{plain}

\bibliography{icooolps_2006_WS_reader}

\end{document}